\documentclass[a4paper,epsfig,prb]{revtex4}
\usepackage{graphicx}
\usepackage{amssymb}
\usepackage{amsmath}

\usepackage{color}

\begin{document}
\title{Electronic Structure and Magnetic Properties of Iridate
  Superlattice SrIrO$_3$/SrTiO$_3$}

\author{Kang-Hwan Kim$^{1*}$, Heung-Sik Kim$^{1*}$\footnotetext{* These two authors contributed equally.}, 
and Myung Joon Han$^{1,2}$}
\address{
$^1$Department of Physics, Korea Advanced Institute of Science and Technology, Daejeon, 305-701, Korea}
\address{
$^1$KAIST Institute of NanoCentury, Korea Advanced Institute of Science and Technology, Daejeon, 305-701, Korea}

\email{mj.han@kaist.ac.kr}

\begin{abstract}
Motivated by an experimental report of iridate superlattices, we
performed first-principle electronic structure calculations for
SrIrO$_3$/SrTiO$_3$. Heterostructuring causes SrIrO$_3$ to become
Sr$_2$IrO$_4$-like, and the system has the well-defined $j_{\rm eff}$ =
1/2 states near the Fermi level as well as canted
antiferromagnetic order within the quasi-two-dimensional IrO$_2$
plane. In response to a larger tensile
strain, the band gap is increased due to the resulting increase in bond length and
the bandwidth reduction. 
The ground state magnetic properties are discussed in comparison to the  metastable 
collinear antiferromagnetic state.
Our work sheds new light for understanding the recent experimental results on 
the iridate heterostructures. 
\end{abstract}

\maketitle

%%%%%%%%%%%%%%%%%%%%%%%%%%%%%%%%%%%%%%%%%%%%%%%%%%%%%%%%%%%%%%%%%%%%%%%%%%%%%%%%%%%%%%%%%%%%%%%

\section{Introduction}
Recently atomic spin-orbit coupling (SOC) has
become a central issue in condensed matter physics \cite{Datta90,Qi09, Mattheiss76}.  
In strong SOC, spin and orbital degrees of freedom are entangled, 
and this characteristic often plays a key role in the emergence of new material properties.
Iridium oxide compounds are of special interest due to the fact that
other atomic energy scales, such as $U$ and $J$, happen to be
in a size comparable to SOC in this class of materials.  Owing to the
cooperation between those energy scales, novel low-energy states,
designated by $j_{\rm eff}$ (the effective total angular
momentum)\cite{BJKim08, BJKim09}, and exotic quantum ground states
can be realized \cite{Krempa13, Kane05, Chang13, Shitade09}.

The Ruddlesden-Popper (RP) series of iridates,
Sr$_{n+1}$Ir$_{n}$O$_{3n+1}$, are known to exhibit quite different
features depending on $n$.  Ideally, by making the lower dimensional
forms, one can change and even control their properties. Several
recent reports on the iridate thin films
\cite{Ramesh13,Nichols13,Nichols13a,Jenderka13} provide a good
playground for investigation from this perspective. Also, SrIrO$_3$/SrTiO$_3$ (SIO/STO)
superlattice (SL) seems to have been synthesized successfully
\cite{Takagi12}. According to the recent report by Matsuno {\it et al.}, 
controlled and systematic changes are found in
[SIO]$_m$/[STO]$_1$ as a function of SIO layer thickness,
$m$. These experiments open a new stage in the study of
iridate physics by enabling control of their dimensionality and
therefore of their material characteristics.

One of the most interesting questions in these SLs may be how close
the thinnest SL can be to Sr$_2$IrO$_4$ because SIO and Sr$_2$IrO$_4$ are
the two end members of the RP series. If the property of {\it $m$=$1$} 
SL is similar to those of Sr$_2$IrO$_4$, the physics of the whole RP
series can in principle be accessed in the SL by changing SIO
thickness.  Further, epitaxial strain provides an extra dimension
for controlling their properties.

With this motivation, we performed density functional theory
calculations for this SIO/STO. It was found that this
form of SL actually exhibits quite similar characteristics to
Sr$_2$IrO$_4$ and is notably different from the other RP iridates, such as
Sr$_3$Ir$_2$O$_7$. It has an insulating ground state and canted
antiferromagnetic (AF) ordering of $j_{\rm eff}$ = 1/2 moments.  The
electronic and magnetic properties are systematically changed by
strain. Our results shed new light for understanding the novel
correlated spin-orbital physics of iridate compounds.

\section{Computational details}
First-principles electronic structure calculations were performed
by OPENMX code \cite{MJHan06, openmx}, which is based on the
linear combination of pseudoatomic orbitals method \cite{ozaki03}.
The Perdew-Burke-Ernzerhof (PBE) exchange-correlation functional
\cite{Perdew96} was adopted.  The energy cutoff of 300 Ry and the
Monkhorst-Pack $k$-meshes of 7 $\times$ 7 $\times$ 5 in the first
Brillouin zone were used for the real and momentum space integrations,
respectively.  SOC was treated within a fully relativistic
$j$-dependent pseudopotential scheme in the non-collinear methodology
\cite{openmx}.  Electronic correlations were taken into account with the
DFT+$U$ formalism \cite{Dudarev}. We used the effective
on-site Coulomb interaction parameter of $U_{\rm eff}\equiv U-J=2.0$
eV for the Ir $5d$ orbitals \cite{Arita12, Foyevtsova13}.  The
optimized c-axis lattice constant and the internal coordinates are
compared well with the result by Vienna {\it ab-initio} Simulation
Package \cite{VASP1,VASP2}.  The energy criteria was $10^{-3}$ eV for
the structural optimization.

\begin{figure}
  \centering
  \includegraphics[width=0.7\textwidth]{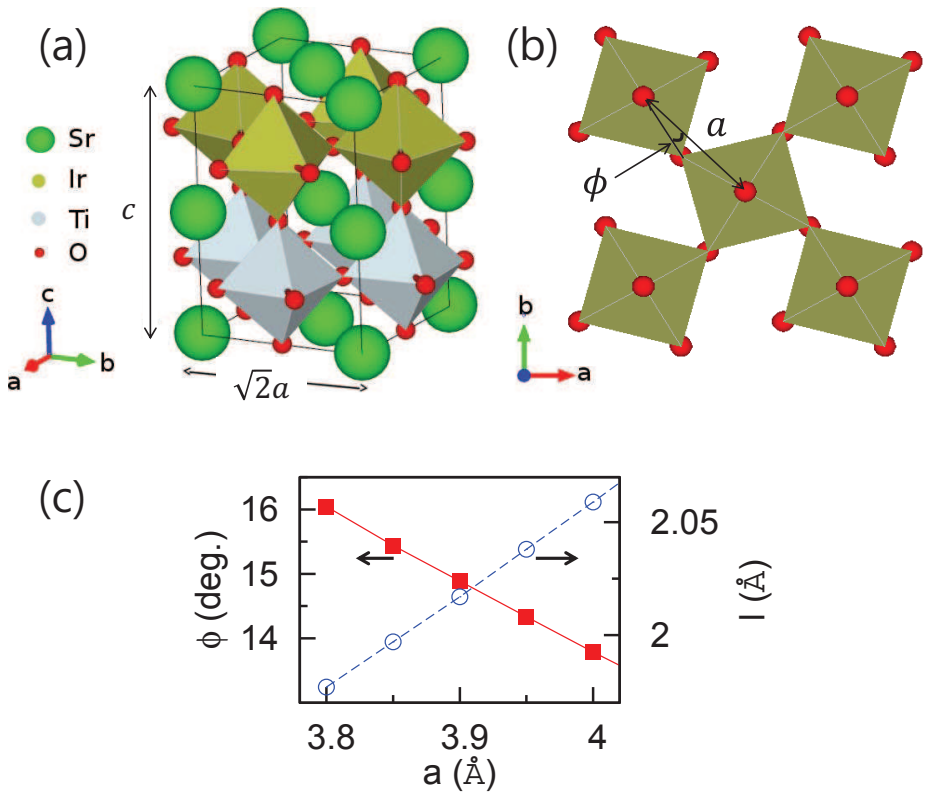}
  \caption{(a) The unicell structure used in this study. (b) Top
    view of the optimized geometry of IrO$_6$ cages where $a$ and
    $\phi$ refer to the distance between the nearest Ir atoms and the
    Ir-O-Ir bond angle, respectively. (c) The calculated structure
    parameters as a function of in-plain lattice constants,
    $a$. $\phi$ and $l$ denote the octahedral rotation angle and the
    Ir-O bond length, respectively.}
  \label{fig:structure}
\end{figure}

\section{Results and discussion}
\subsection{Structural properties}
Fig.~\ref{fig:structure}(a) shows the
unitcell structure used in this study. As in Sr$_2$IrO$_4$, this SL
has TiO$_2$ inter-layers which are electronically inactive ($d^0$
configuration) and effectively suppress hopping between the
neighboring IrO$_2$ layers. Therefore our system becomes
quasi-two-dimensional, as is the case in Sr$_2$IrO$_4$.  It was also found
that the larger IrO$_6$ octahedra have antidistortive rotations
along the c-axis in order to fit into the smaller SrO$_6$ cage (with the
rotation angle $\phi$) as depicted in Fig.~\ref{fig:structure}(b).  This
produces an environment for the electronic behavior similar to the Ir ions
in Sr$_2$IrO$_4$.

To simulate the epitaxial strain, the in-plain lattice constants were
varied from 3.80 to 4.00\AA.  Note that the equilibrium in-plane
lattice constant of Sr$_2$IrO$_4$ is $\sim$3.88 \AA~ \cite{Nichols13,
  Crawford94}.  If we assume the IrO$_6$ octahedra are rigid ({\it
  i.e.,} no change in the Ir-O bond lengths), it is natural to expect
the enhancement (reduction) of the octahedral rotation for the
compressive (tensile) strain.  In fact, our result qualitatively
follows such a tendency in the rotation pattern. Simultaneously,
however, we also found that the IrO$_6$ octahedra undergo
significant tetragonal distortions, as summarized in
Fig.~\ref{fig:structure}(c).  The in-plane Ir-O bond length is changed
 by  $\sim\pm$2.0\% as $a$ is increased from 3.90 \AA~ to 4.00 \AA
(tensile strain) or decreased down to 3.80 \AA (compressive strain).

\subsection{Electronic properties without $U_{\rm eff}$}
		
\begin{figure}
\centering
\includegraphics[width=0.7\textwidth]{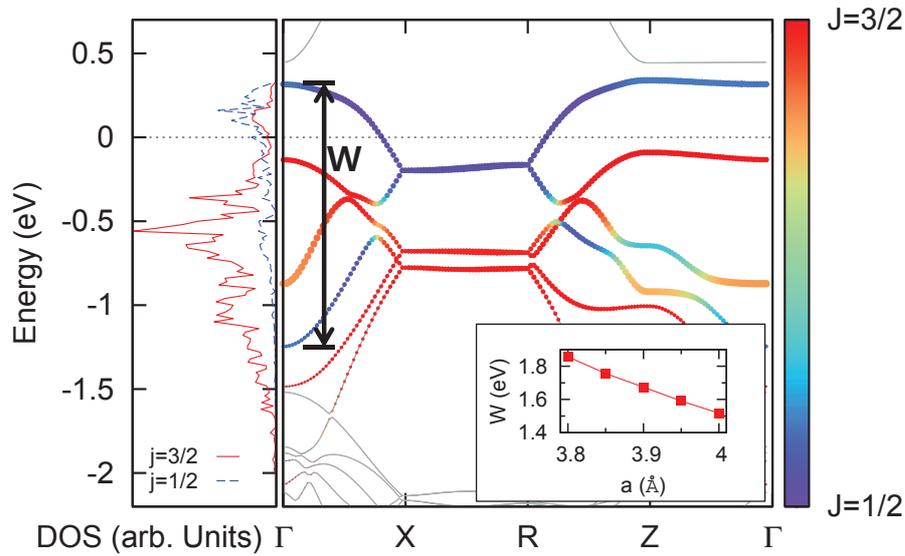}
  \caption{The calculated band structure with SOC ($U_{\rm eff}$=0 and
    with a=3.95 \AA). The size and color of the point represent the amount of
    $t_{2g}$ and $j_{\textrm{eff}}$-character, respectively.  
    (Inset) The dependence of the
    calculated bandwidths on the in-plain lattice constant.}
  \label{fig:soc}
\end{figure}

Fig.~\ref{fig:soc} presents the calculated band structure with SOC
($U_{\rm eff}$=0).  
The size and color of each band point represents
the portions of $t_{\rm 2g}$ and $j_{\rm eff}$ states, respectively. 
As expected, the $t_{\rm 2g}$ characters prevail around the
Fermi energy. The threefold degenerate $t_{\rm 2g}$ states carry 
the effective orbital angular momentum $l_{\rm eff}=1$. 
Assuming that the SOC of the Ir atom dominates over the
other energy scales, such as bandwidths and tetragonal crystal fields,
the $(l_{\rm eff}=1) \oplus (s=1/2)$ states are split into an upper $j_{\rm eff}$=1/2
doublet and a lower $j_{\rm eff}$=3/2 quartet \cite{BJKim08}, 
where the $j_{\rm eff}=1/2$ and $3/2$ states are defined as
\begin{eqnarray}
\vert j_{\rm eff}=\frac{1}{2},\pm\frac{1}{2} \rangle_z 
&\equiv& 
\mp \sqrt{\frac{1}{3}} \vert d_{xy} \rangle \vert \uparrow\downarrow \rangle_z
\mp \sqrt{\frac{2}{3}} \left[ \frac{\pm \vert d_{yz} \rangle + i\vert d_{xz} \rangle}{\sqrt{2}} \right] \vert \downarrow\uparrow \rangle_z \nonumber \\
&\equiv&
\pm \sqrt{\frac{1}{3}} \vert 0, \uparrow\downarrow \rangle_z \mp \sqrt{\frac{2}{3}} \vert \pm1, \downarrow\uparrow \rangle_z \nonumber \\
\vert j_{\rm eff}=\frac{3}{2},\pm\frac{1}{2} \rangle_z 
&\equiv& 
\sqrt{\frac{2}{3}} \vert 0, \uparrow\downarrow \rangle_z + \sqrt{\frac{1}{3}} \vert \pm1, \downarrow\uparrow \rangle_z \nonumber \\
\vert j_{\rm eff}=\frac{3}{2},\pm\frac{3}{2} \rangle_z 
&\equiv& 
\vert \pm1, \uparrow\downarrow \rangle_z 
\label{eq:jeff}
\end{eqnarray}
where $\{\uparrow, \downarrow\}$ and $\{0,\pm1\}$ denote
the indices for the spin and the angular momentum eigenstates $\vert s=1/2; \pm1/2 \rangle_z$ and $\vert l_{\rm eff}=1; 0,\pm1 \rangle_z$,
respectively, and the subscript $z$ means that the momenta are quantized along the octahedral $z$-direction 
(perpendicular to the Ir plane).
Although the splitting between the $j_{\rm eff}=1/2$ and $3/2$ states is not perfect due to the
sizeable effect from the bandwidth and the confinement in the SL, the $j_{\rm eff}=1/2$ 
characters in the low-energy states near the Fermi level are quite clearly noticed (blue colors in
Fig.~\ref{fig:soc}). It is found that the $j_{\rm eff}$-characterization of the bands is valid in the whole range of
strains we considered in this study.  There are some minor
differences between the band structure of SL in Fig.~\ref{fig:soc}
and the bulk
Sr$_2$IrO$_4$: In Fig.~\ref{fig:soc}, the position
of the valence band top at X point is lower by
$\sim$0.1eV than that at $\Gamma$, while they are nearly at the
same energy in Sr$_2$IrO$_4$ \cite{BJKim08,Zhang13}. The same feature is
also observed in the $U>0$ calculations (Fig.~\ref{fig:SO+U}(a)). The
presence of the additional tetragonal crystal field, induced by the
interface effect and the strain, yields some small changes in the
relative position of the $j_{\rm eff}$=1/2 and 3/2 bands compared to
the bulk Sr$_2$IrO$_4$.

The bandwidth $W$ of the low-energy $j_{\rm eff}$=1/2 states,
which are located near the Fermi level, is of key importance
in understanding the electronic property of RP iridates
\cite{SJMoon09}.  One way of defining $W$ is at the $\Gamma$ point as
shown in Fig.~\ref{fig:soc}.  The inset of Fig.~ \ref{fig:soc}
presents the change of $W$ as a function of strain.  It is clearly
seen that the bandwidth $W$ is decreased as larger tensile strain
is applied. This result can be surprising if one tries to understand
the system based on the `rigid IrO$_6$ octahedron' picture (see, for
example, the discussion in Nichols {\it et al.}\cite{Nichols13}). It is
therefore important to note that the Ir-O bond lengths are increased
(decreased) under tensile (compressive) strain as mentioned above.
Since enlarged bond lengths generally reduce hopping, the
reduction of the bandwidth by the tensile strain is attributed to the
bond length changes.

\subsection{Electronic and magnetic properties with $U_{\rm eff}$}

\begin{figure}
\centering
\includegraphics[width=0.7\textwidth]{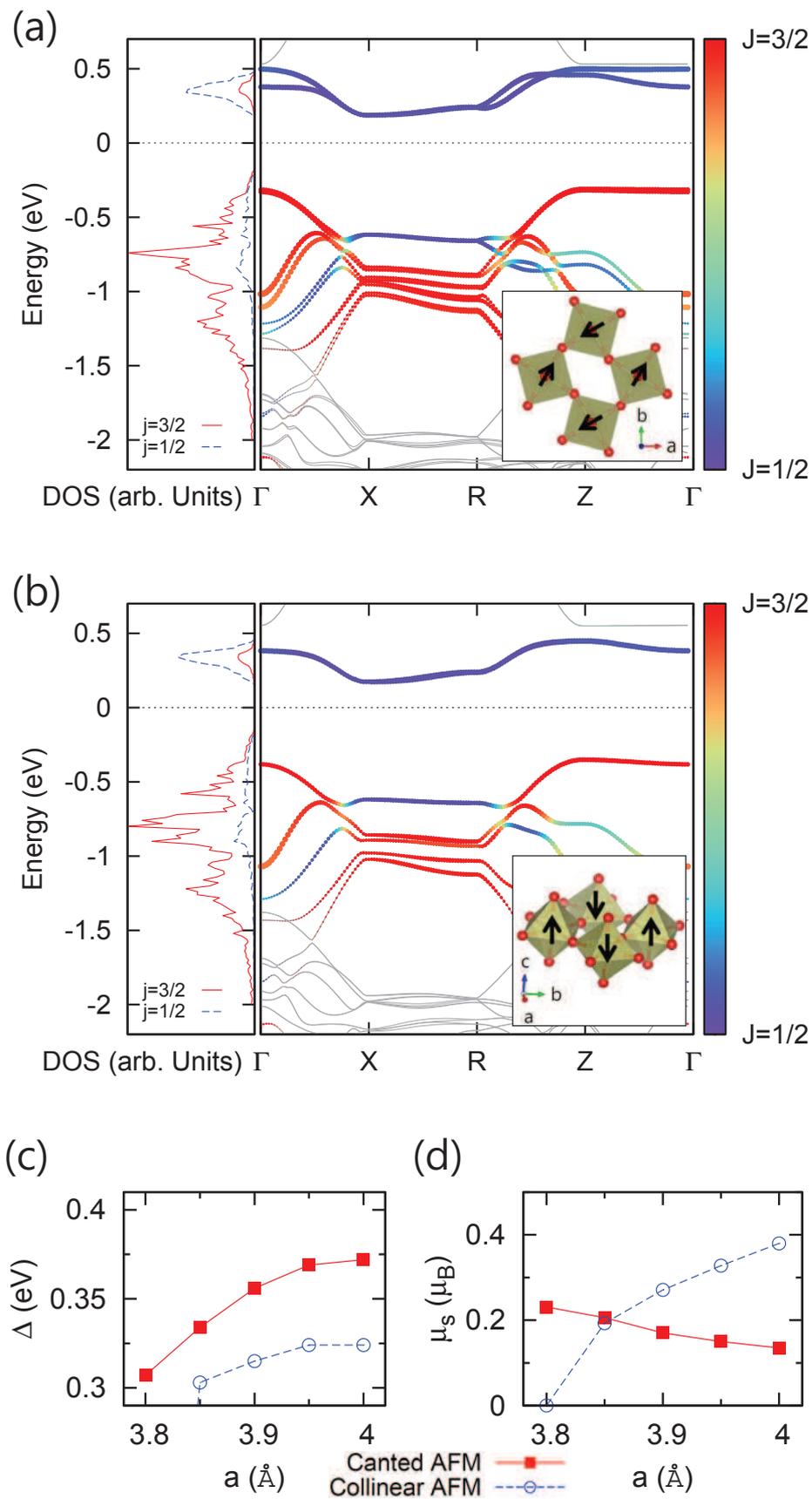}
  \caption
  {The calculatd band structure calculated with SOC and $U_{\rm eff}$=2 eV
    ($a=3.95$ \AA~) for (a) the canted AF and (b) collinear AF
    order. The calculated (c) band gap $\Delta$ and (d) spin magnetic
    moment as a function of the in-plain lattice constant.  }
\label{fig:SO+U}
\end{figure}

Fig.~\ref{fig:SO+U}(a)-(d) summarizes the calculation results with SOC
and $U_{\rm eff}$=2 eV. Two different magnetic configurations have been
considered; the canted AF spin order (the ground state configuration
of Sr$_2$IrO$_4$; see the inset of Fig.~\ref{fig:SO+U}(a)) and the
collinear AF order (the ground state of Sr$_3$Ir$_2$I$_7$; see the inset of
Fig.~\ref{fig:SO+U}(b)).  In the canted AF state, the Ir moments lie within
the $ab$-plane and their angles follow the tilting angles of
IrO$_6$ octahedra. In the collinear AF state, the moments are parallel
to the c-axis. It is found that, as in Sr$_2$IrO$_4$, the canted AF
state is more stable than the collinear AF state by about 3 meV/Ir
over the strain range considered in this study.

In both magnetic configurations, the electronic correlation is found
to play the key role in opening the gap and stabilizing the AF order
as well as the dominant $j_{\rm eff}$=1/2 character in the upper
Hubbard bands. Similar features were also observed in the previous
study of Sr$_2$IrO$_4$ \cite{HJin09}. The calculated band gap of
the canted and the collinear AF phase at $a=3.90$\AA~, 0.35 and 0.31 eV, respectively,
%% $\sim$0.35 eV at $a=3.90$\AA~ 
is comparable to the optical gap of
$\sim$0.5 for Sr$_2$IrO$_4$ bulk and thin films 
\cite{SJMoon09,Nichols13}, considering a little ambiguity in the $U_{\rm eff}$ 
parameter. Note that, when a larger (tensile) strain is applied,
the band gap is markedly increased as shown in Fig.~\ref{fig:SO+U}(c),
which is consistent with the decreasing trend of $W$ (see
Fig.~\ref{fig:soc}).  Importantly, this trend is in a good agreement
with recent optical spectroscopy data by Nichols {\it et al.} for
the Sr$_2$IrO$_4$ thin film under various strain conditions in which the
enhancement of the optical gap upon tensile strain was observed
\cite{Nichols13}.  This feature is hard to understand from the
point of view of the conventional rigid IrO$_6$ picture as discussed
in Nichols {\it et al.}\cite{Nichols13}, whereas our result based on
the tetragonal distortion of the IrO$_6$ octahedra
provides a natural explanation.

Although the collinear AF phase is less stable than the canted AF, the
data from this configuration also provides useful information.  First,
we note that the gap size in the collinear AF is smaller than that of
the canted AF state over the entire range of strain. It is interesting
to note that the gap is closed at $a=3.80$\AA~ and the spin moment
vanishes simultaneously. 
Since [SIO]$_1$/[STO]$_1$ SL has considerable similarities in the electronic and
magnetic properties with Sr$_2$IrO$_4$, 
this might provide clues for understanding the
peculiar magnetoresistance behavior recently reported in Sr$_2$IrO$_4$
\cite{MGe}.  The experiment by Ge {\it et al.}\cite{MGe} seems to indicate a
spin-flop-like transition upon application of the external fields parallel to the
$c$-axis.  This implies a possible switching of the magnetic
configuration by external fields and a significant change of the
electronic structure in Sr$_2$IrO$_4$.
Our results also suggest a
possible insulator-to-metal transition driven by external
magnetic field in the epitaxially strained SIO superlattices and ultrathin films.

Fig.~\ref{fig:SO+U}(d) shows the change of spin moments with respect to strain. 
While the spin moment is reduced by increasing tensile strains in the canted AF phase, 
it is enhanced in the collinear AF state.
This feature can be understood by considering the strain-induced tetragonal crystal fields
and the low-energy states which have deviated from the ideal $j_{\rm eff}$=1/2 as follows.
Let us consider the $t_{\rm 2g}$ states with the dominating SOC, $\lambda$, and 
the nonvanishing tetragonal fields, $\Delta_{\rm t}$. The effective atomic Hamiltonian is
\begin{equation}
\mathcal{H}_{\rm eff} = \lambda \mathbf{l}_{\rm eff} \cdot \mathbf{s} + \Delta_{\rm t}l_{\rm eff,z}^2,
\label{eq:Heff}
\end{equation}
where $\lambda<0$ for the $t_{\rm 2g}$ states.
Among the six eigenstates of Eq. (\ref{eq:Heff}), 
the upper twofold-degenerate states are given by \cite{Jackeli09}
\begin{equation}
\vert \widetilde{\pm}\rangle_z = \pm \sin \theta \vert 0,\uparrow\downarrow \rangle_z 
\mp \cos \theta \vert \pm1, \downarrow\uparrow \rangle_z,
\label{eq:isospin}
\end{equation}
or, when quantized along the in-plane (say, the octahedral $x$-direction),
\begin{equation}
\vert \widetilde{\pm}\rangle_x = 1/\sqrt{2}(\vert \widetilde{+} \rangle_z \pm \vert \widetilde{-} \rangle_z). 
\label{eq:isospin_z}
\end{equation}
The orbital angle $\theta$ is defined as $ \tan (2\theta) = 2\sqrt{2} \lambda /(\lambda - 2 \Delta_{\rm t})$, 
and in the perfect cubic symmetry ($\Delta_{\rm t} = 0$, $\theta_{\rm cubic} \simeq 0.2 \pi$) Eq. (\ref{eq:isospin}) reduces to the ideal
$j_{\rm eff}=1/2$ states defined in Eq. (\ref{eq:jeff}).
From Eq. (\ref{eq:Heff}) one can notice that the presence of the tensile strain or the confinement effect, 
which lowers the energy of the $d_{xy}$ state ($\Delta_{\rm t} > 0$), corresponds to the smaller $\theta$.
The strain dependence of the ordered spin moment in the collinear and the canted AF phases
can be estimated from Eq. (\ref{eq:isospin}) and (\ref{eq:isospin_z}), respectively.
For the collinear AF, $\vert \widetilde{\pm} \rangle_z$ 
forms the lower/upper Hubbard bands to yield the local spin and orbital moments parallel to the $z$-axis on the Ir sites. The expectation value of $s_z$ is
\begin{equation}
\langle \widetilde{\pm} \vert s_z \vert \widetilde{\pm} \rangle_z   = \mp 1/2 \cos(2 \theta).
\label{eq:collinear_spin}
\end{equation}
In the canted AF phase, on the other hand, the moments are  parallel to the 
octahedral $x$-axis, and the expectation value is
%% Similarly, the same quantity along the in-plane direction, say $x$, can be written down
%% for $\vert \widetilde{\pm}\rangle_x$ quantized along the octahedral $x$-direction, which reads
\begin{equation}
\langle \widetilde{\pm} \vert s_x \vert \widetilde{\pm} \rangle_x    =  \mp 1/4 (1-\cos(2 \theta)).
\label{eq:canted_spin}
\end{equation}
It is clear that  $\lvert \langle \widetilde{\pm} \vert s_z \vert \widetilde{\pm} \rangle_z \rvert$ and 
$\lvert \langle \widetilde{\pm} \vert s_x \vert \widetilde{\pm} \rangle_x \rvert$ behave in different ways
with respect to $\Delta_{\rm t}$; with larger tensile strains, 
$\lvert \langle \widetilde{\pm} \vert s_z \vert \widetilde{\pm} \rangle_z \rvert$ is enhanced 
and $\lvert \langle \widetilde{\pm} \vert s_x \vert \widetilde{\pm} \rangle_x \rvert$ is reduced.
This is consistent with the results of Fig. \ref{fig:SO+U}(d)
in the region $\theta < \pi/4$. Since the confinement effect in the SL introduces a
significant amount of $\Delta_{\rm t} > 0$, 
the $\theta$ is smaller than $\theta_{\rm cubic} < \pi/4$ over the entire range of strains.

\section{Concluding remarks}
We have investigated the structural, electronic, and
magnetic properties of SIO/STO SL. 
The results clearly demonstrate the similarity between [SIO]$_1$ /[STO]$_1$ 
SL and bulk Sr$_2$IrO$_4$, as expected from
their structural similarities.
By adding extra dimensions of controllability, the SL form of
 iridates can provide interesting new information for
 understanding the $5d$ transition metal oxides. It can be an important next step to
 investigate [SIO]$_2$/[STO]$_{n\geq 1}$ in comparison with
 Sr$_3$Ir$_2$O$_7$ \cite{Zhang13, JWKim12}.  As the
 SIO-layer thickness is increased from the ultrathin limit to the bulk
 regime, the
 electronic structure can evolve from the insulating to the nodal
 semimetallic phase which also deserves intensive future study  \cite{Takagi12,SJMoon09,Kee12}.

\section*{Acknowledgements}
We thank Sung Seok (Ambrose) Seo, Hosub Jin, and Jaejun Yu for helpful
discussion.  This research was supported by Basic Science Research
Program through the National Research Foundation of Korea(NRF) funded
by the Ministry of Education(Grant No. 2013R1A6A3A01064947).
Computational resources were provided by the National Institute of
Supercomputing and Networking/Korea Institute of Science and
Technology Information with supercomputing resources including
technical support (Grant No. KSC-2013-C2-024).

\end{document}